# Some new models of anisotropic relativistic stars in linear and quadratic regime

Manuel Malaver[1] , Rajan Iyer[2]

[1]Maritime University of the Caribbean, Department of Basic Sciences, Catia la Mar, Venezuela.
 Email: **mmf.umc@gmail.com**

[2]Environmental Materials Theoretical Physicist, Department of Physical Mathematics Sciences Engineering
  Project Technologies, Engineeringinc International Operational Teknet Earth Global, Tempe, Arizona,
  United States of America
  Email: engginc@msn.com

**Abstract:** Taking local anisotropy into consideration, in this paper, some new analytical models of relativistic anisotropic charged quark stars in linear and quadratic regime have been developed. The Einstein-Maxwell field equations have been solved with a particular form of metric potential and electric field intensity. The plots generated show that physical variables such as metric potentials, radial pressure, energy density, charge density, anisotropy, radial speed sound are consistent with realistic stellar models. We obtained some models consistent with stellar objects as 4U 1538-52, Cyg X-2, OGLE-TR-122b and red dwarf Proxima Centauri.

## 1.- Introduction

The Einstein´s theory of general relativity (GR) is considered as a fundamental theory to understand the behavior and structure of stellar massive objects as neutron stars, quasars, pulsars and white dwarfs [1,2].The first theoretical study in GR is carried out by Karl Schwarzschild [3] in 1916 with the discovery of a universal vacuum exterior solution. Also Einstein makes important contributions to the research on the structure of the universe producing relevant advances in the fields of astrophysics and cosmology [4].

   For many decades, the star interior was considered made of perfect fluid where are equal the radial ($p_r$) and tangential ($p_t$) pressures and this leads to the isotropic local condition $p_r = p_t$ [5] but the theoretical studies of Ruderman [6] and Canuto [7] on realistic star models suggest that the nuclear matter can present anisotropy in high density ranges ( $\rho > 10^{15}$ g cm$^3$). In the massive objects the radial pressure differs from tangential and the anisotropy in the pressures can be induced by a solid core, phase transitions, presence of magnetic field and pion condensation [8,9]. Usov [10] suggest that strong electric field may also cause pressure anisotropy. In the pioneering work of Bowers and Liang [11] consider an anisotropic fluid model and conclude that anisotropy influence in the maximum equilibrium mass and surface redshift. Also Dev and Gleiser [8] have presented new exact solutions for distribution of matter with tangential pressures and uniform energy density and propose an equation which relates the radial and tangential pressures in order to integrate analytical the Einstein field equations. Authors as Cosenza et al. [12], Bayin [13], Krori et al. [14], Herrera and Ponce de León [15], Ponce de León [16, 17], Bondi [18], Herrera and Santos [19], Herrera et al. [20, 21 ], Dev and Gleiser [8, 22], Ivanov [23,24], Mak and Harko [25, 26], Mak et al. [27], Viaggiu [28], Malaver [29-34], Pant et al. [35] have studied the effect of anisotropy on the physical properties of a distribution of matter in relativistic compact objects.

The possibility of the existence of strange quark stars in hydrostatic equilibrium was first speculated for Itoh [36] in a seminal treatment. In the hypothesis of the strange quark matter, the quark matter consists of equal number of up, down quarks and the strange quarks can be considered as the absolute ground state for the confined state of hadrons [37]. It is know that strange stars form during the collapse of the core of a massive object after a supernova explosion so this kind of stars are not part of equilibrium configurations as neutron stars and white dwarf.

An important distinction between strange stars and conventional neutron stars is that the strange stars are self-bound by the strong interaction, gravity just make them massive, whereas neutron stars are bound by gravity. This allows a strange star to rotate faster than would be possible for a neutron star. The most fascinating distinction between a strange star and a normal neutron star is the surface electric fields associated with it. Strange stars possess ultra-strong electric fields on their surfaces, which, for ordinary strange matter, is around $10^{18}$ V/cm and $10^{20}$ V/cm for color superconducting strange matter [38,39]. The influence of energy densities of ultra-high electric fields on the bulk properties of compact stars was explored in [39–45]. It also has been shown that electric fields of this magnitude, generated by charge distributions located near the surfaces of strange stars, increase the stellar mass by up to 30% depending on the strength of the electric field. These features may allow one to observationally distinguish strange stars from neutron stars.

Many researches as Ponce de León [16], Patel and Vaidya [46], Tikekar and Thomas [47], Mak and Harko [48], Chaisi and Maharaj [49–51], Maharaj and Chaisi [52,53], Sharma and Maharaj [54], Varela et al. [55], Komathiraj and Maharaj [56], Takisa and Maharaj [57–59], Maharaj and Takisa [60], Feroze and Siddique [61,62], Feroze [63], Feroze and Tariq [64], Bhar et al. [65], Murad [66] and Malaver [67,68] have developed mathematically exact analytical models of strange stars with a linear equation of state based on MIT bag model and with different metric potentials. We would too like in pointing out that novel paradigm PHYSICS necessitates knowing what is going on inside these galaxies and stars especially that are anisotropic may hold clue to the quantum nature [69] that ansatz new paradigm PHYSICS has been developed to formalisms that have sound logic with assumption-free theoretical advancements that has ability to be gaged from the quantum to astrophysics regime metrically [70-73]. Also, Bhar and Murad [74] consider a Chaplygin equation of state with locally anisotropic matter distribution using a particular type of metric function. Pure physical time transforms have been ansatz developed, that have not been available before in literature, that has capability to prove universe is like a black box, however, observable parameters are derivable in terms of density matrix and four-vector time fields matrix gaged to switch mode four-vector-ket matrix [75]. Currently, findings of the James Webb telescope discovered six galaxies that appeared between 500 and 700 million years after the Big Bang that essentially seem to contradict the present understanding of how the first galaxies were formed [76]. Hence, quantum astrophysics theoretical to experimental observable measurement methodologies with reworking formalism from Helmholtz decomposition then algorithmically gaging to fibrational discontinuum energy fields with a four-vector time fields to four-vector-ket matrix and four element wavefunction point-to-point bra matrix together displaying astrophysical observable signal/noise matrix will be experimentally required to get to the bottom of this new PHYSICS [70-76].

As stated the principal motivation of this work is to develop some new analytical relativistic stellar models by obtaining closed-form solutions of Einstein-Maxwell field equations in presence of electric field and with anisotropy in linear and quadratic regime. We have used the metric potential proposed by Thirukkanesh and Ragel [77] and generalized for Malaver [29] or Thirukkanesh-Ragel-Malaver ansatz. This paper is structured as follows: The next section, Sect. 2, we show the solution of Einstein-Maxwell field equations of anisotropic fluid and derives the pressure and density relation. In Sect. 3 we present a new classes of models for compact objects. In Sect. 4 physical acceptability conditions are discussed. The new models obtained are physically analyzed in Sect. 5. The conclusions of the results obtained are shown in Sect. 6.

## 2. Einstein-Maxwell Field Equations

We consider a spherically symmetric, static and homogeneous spacetime. In Schwarzschild coordinates the metric is given by

$$ds^2 = -e^{2\nu(r)}dt^2 + e^{2\lambda(r)}dr^2 + r^2(d\theta^2 + \sin^2\theta d\varphi^2) \tag{1}$$

where $\nu(r)$ and $\lambda(r)$ are two arbitrary functions.

The Einstein field equations for the charged anisotropic matter are given by

$$\frac{1}{r^2}\left(1-e^{-2\lambda}\right) + \frac{2\lambda'}{r}e^{-2\lambda} = \rho + \frac{1}{2}E^2 \tag{2}$$

$$-\frac{1}{r^2}\left(1-e^{-2\lambda}\right) + \frac{2\nu'}{r}e^{-2\lambda} = p_r - \frac{1}{2}E^2 \tag{3}$$

$$e^{-2\lambda}\left(v''+v'^2+\frac{v'}{r}-v'\lambda'-\frac{\lambda'}{r}\right)=p_t+\frac{1}{2}E^2 \qquad (4)$$

$$\sigma=\frac{1}{r^2}e^{-\lambda}(r^2 E)' \qquad (5)$$

where $\rho$ is the energy density, $p_r$ is the radial pressure, $E$ is electric field intensity, $p_t$ is the tangential pressure and primes denote differentiations with respect to r. Using the transformations, $x=Cr^2$, $Z(x)=e^{-2\lambda(r)}$ and $A^2 y^2(x)=e^{2v(r)}$ with arbitrary constants A and c>0, suggested by Durgapal and Bannerji [78], the Einstein field equations can be written as

$$\frac{1-Z}{x}-2\dot{Z}=\frac{\rho}{C}+\frac{E^2}{2C} \qquad (6)$$

$$4Z\frac{\dot{y}}{y}-\frac{1-Z}{x}=\frac{p_r}{C}-\frac{E^2}{2C} \qquad (7)$$

$$4xZ\frac{\ddot{y}}{y}+(4Z+2x\dot{Z})\frac{\dot{y}}{y}+\dot{Z}=\frac{p_t}{C}+\frac{E^2}{2C} \qquad (8)$$

$$p_t=p_r+\Delta \qquad (9)$$

$$\frac{\Delta}{C}=4xZ\frac{\ddot{y}}{y}+\dot{Z}\left(1+2x\frac{\dot{y}}{y}\right)+\frac{1-Z}{x}-\frac{E^2}{C} \qquad (10)$$

$$\sigma^2=\frac{4cZ}{x}(x\dot{E}+E)^2 \qquad (11)$$

$\sigma$ is the charge density, $\Delta = p_t - p_r$ is the anisotropic factor and dots denote differentiation with respect to x. With the transformations of [78], the mass within a radius r of the sphere takes the form

$$m(x) = \frac{1}{4C^{3/2}} \int_0^x \sqrt{x}(\rho^* + E^2) dx \tag{12}$$

where $\rho^* = \left(\frac{1-Z}{x} - 2\dot{Z}\right) C$

The interior metric (1) with the charged matter distribution should match the exterior spacetime described by the Reissner-Nordstrom metric:

$$ds^2 = -\left(1 - \frac{2M}{r} + \frac{Q^2}{r^2}\right) dt^2 + \left(1 - \frac{2M}{r} + \frac{Q^2}{r^2}\right)^{-1} dr^2 + r^2(d\theta^2 + \sin^2\theta d\varphi^2) \tag{13}$$

where the total mass and the total charge of the star are denoted by $M$ and $q^2$, respectively. The junction conditions at the stellar surface are obtained by matching the first and the second fundamental forms for the interior metric (1) and the exterior metric (13).

In this paper, we imposed the following equations of state, linear and quadratic, respectively, relating the radial pressure to the energy density, where $m$ is a positive constant

$$p_r = m\rho \quad \text{and} \quad p_r = m\rho^2 \tag{14}$$

## 3. Classes of Models for compact objects

In order to solve the Einstein-Maxwell field equations, in this work we have chosen the metric potential proposed by Thirukanesh and Ragel [77] and Malaver [29] which has the form $Z(x) = (1 - ax)^n$, where $a$ is a real constant and $n$ is an adjustable parameter and we take the electric field intensity used for Lighuda et al. [79] that can be written as

$$\frac{E^2}{2C} = kxZ(x) = kx(1 - ax)^n \tag{15}$$

with $k$ as an arbitrary real constant. The gravitational potential chosen is regular at the stellar center and well behaved in the interior of the sphere and the electric field is finite at the center of the star and remains continuous in the interior. In this research we have analyzed the particular cases for n=1 for linear equation of state and n=2 for quadratic regime.

For the case n=1, using Z(x) and eq. (15) in eq. (6) we obtain

$$\rho = C[3a - kx(1-ax)] \tag{16}$$

Substituting eq.(16) in eq. (14) for linear regime the radial pressure will be given for

$$p_r = m\rho = mC[3a - kx(1-ax)] \tag{17}$$

With eq. (16) and Z(x) in eq. (11) the charge density is expressed as follows:

$$\sigma^2 = 2C^2 k(3 - 4ax)^2 \tag{18}$$

Using eq. (16) in eq. (12) we obtain for the mass function

$$M(x) = \frac{(5akx^2 - 7kx + 35a)x^{3/2}}{70\sqrt{C}} \tag{19}$$

Substituting (17), (15) and Z(x) in eq. (7) we have

$$\frac{\dot{y}}{y} = \frac{m[3a - kx(1-ax)]}{4(1-ax)} - \frac{kx}{4} + \frac{a}{4(1-ax)} \tag{20}$$

Integrating eq. (20), we obtain

$$y(x) = C_1 (ax - 1)^{\frac{-3m-1}{4}} e^{-\frac{(m+1)kx^2}{8}} \tag{21}$$

Where $C_1$ is the constant of integration

For the metric functions $e^{2\lambda}$ and $e^{2\nu}$ we have

$$e^{2\lambda} = \frac{1}{1-ax} \tag{22}$$

$$e^{2\nu} = A^2 C_1^2 (ax-1)^{\frac{-3m-1}{2}} e^{-\frac{(m+1)kx^2}{4}} \tag{23}$$

and the anisotropy $\Delta$ is given by for

$$\Delta = 4xC(1-ax)\left[\left(\frac{3m+1}{4}\right)^2 \frac{a^2}{(1-ax)^2} + \left(\frac{3m+1}{4}\right)\frac{a^2}{(ax-1)^2} + \frac{a(3m+1)(m+1)kx}{8(ax-1)} - \frac{(m+1)k}{4} + \frac{(m+1)^2 k^2 x^2}{16}\right]$$
$$-aC\left[1+2x\left(-\frac{(3m+1)a}{4(ax-1)} - \frac{(m+1)kx}{4}\right)\right] + aC - kxC(1-ax) \tag{24}$$

With n=2 and quadratic equation of state we found the following expressions for $\rho$, $p_r$, $\sigma^2$, $M(x)$, $e^{2\lambda}$, $e^{2\nu}$ and $\Delta$

$$\rho = C\left[6a - 5a^2 x - kx(1-ax)^2\right] \tag{25}$$

$$p_r = mC\left[6a - 5a^2 x - kx(1-ax)^2\right]^2 \tag{26}$$

$$\sigma^2 = 2C^2 k\left(3 - 8ax + 5a^2 x^2\right)^2 \tag{27}$$

$$M(x) = -\frac{(35a^2 kx^3 - 90akx^2 + 315a^2 x + 63kx - 630a)x^{3/2}}{630\sqrt{C}} \tag{28}$$

$$e^{2\lambda} = \frac{1}{(1-ax)^2} \tag{29}$$

$$e^{2\nu} = A^2 C_2^2 (ax-1)^{\frac{-10Cam-1}{2}} e^{\frac{A*x^6 + Bx^5 + C*x^4 + Dx^3 + Ex^2 + F}{60(ax-1)}} \tag{30}$$

$C_2$ is the constant of integration and

$$\Delta = 4xC(1-ax)^2 \left[ \frac{a^2\left(\frac{1}{4}+\frac{5Cam}{2}\right)^2}{(1-ax)^2} + \left(\frac{1}{4}+\frac{5Cam}{2}\right)\frac{a^2}{(ax-1)^2} + \frac{1}{ax-1}\left( \begin{array}{c} -2a\left(\frac{1}{4}+\frac{5Cam}{2}\right)\left(\frac{6A*x^5+5Bx^4+4C*x^3+3Dx+2Ex}{120(ax-1)}\right) \\ -\frac{a(A*x^6+Bx^5+C*x^4+Dx^3+Ex^2+F)}{120(ax-1)^2} \end{array} \right) \right.$$

$$+ \frac{30A*x^4+20Bx^3+12C*x^2+6Dx+2E}{120(ax-1)} - \frac{(60A*x^5+5Bx^4+4C*x^3+3Dx^2+2Ex)a}{60(ax-1)^2}$$

$$\left. + \frac{(A*x^6+Bx^5+C*x^4+Dx^3+Ex^2+F)a^2}{60(ax-1)^3} + \left( \frac{6A*x^5+5Bx^4+4C*x^3+3Dx^2+2Ex}{120(ax-1)} - \frac{(A*x^6+Bx^5+C*x^4+Dx^3+Ex^2+F)a}{120(ax-1)^2} \right)^2 \right]$$

$$-2aC(1-ax)\left[1+2x\left(-\frac{\left(\frac{1}{4}+\frac{5Cam}{2}\right)a}{ax-1} + \frac{6A*x^5+5Bx^4+4C*x^3+3Dx^2+2Ex}{120(ax-1)} - \frac{a(A*x^6+Bx^5+C*x^4+Dx^3+Ex^2+F)}{60(ax-1)^2}\right)\right]$$

$$+2aC - a^2Cx - kxC(1-ax)^2 \qquad (31)$$

For convenience we have let

$$A* = 6Ca^3k^2m \;,\; B = -21Ca^2k^2m,\; C* = 100Ca^3km + 25Cak^2m \qquad (32)$$

$$D = -(280Ca^2km + 10Ck^2m + 15ak) \;,\; E = 750Ca^3m + 180Cakm + 15k \qquad (33)$$

$$F = -30Cam - 30 \qquad (34)$$

## 4. Conditions of physical acceptability for new models

For a model to be physically acceptable, the following conditions should be satisfied [29, 80]:

(i) The metric potentials $e^{2\lambda}$ and $e^{2\nu}$ assume finite values throughout the stellar interior and are singularity-free at the center $r=0$.

(ii) The energy density $\rho$ should be positive and a decreasing function inside the star.

(iii) The radial pressure also should be positive and a decreasing function of radial parameter.

(iv) The radial pressure and density gradients $dp_r/dr \leq 0$ and $d\rho/dr \leq 0$ for $0 \leq r \leq R$.

(v) The anisotropy is zero at the center $r=0$, i.e. $\Delta(r=0) = 0$.

(vi) Any physically acceptable solution must satisfy the causality condition where the radial speed of sound $v_{sr}^2$ should be less than speed of light throughout the model, i.e.

$$0 \leq v_{sr}^2 = \frac{dp_r}{d\rho} \leq 1.$$

(vii) The interior solution should match with the exterior of the Reissner-Nordstrom spacetime, for which the metric is given by

$$ds^2 = -\left(1 - \frac{2M}{r} + \frac{Q^2}{r^2}\right)dt^2 + \left(1 - \frac{2M}{r} + \frac{Q^2}{r^2}\right)^{-1} dr^2 + r^2 d\theta^2 + r^2 \sin^2\theta d\varphi^2$$

through the boundary $r=R$ where $M$ and $Q$ are the total mass and the total charge of the star, respectively.

The conditions (ii) and (iv) imply that the energy density must reach a maximum at the centre and decreasing towards the surface of the sphere.

## 5. Physical analysis of the new models

With $n=1$ in linear regime the metric potentials $e^{2\lambda}$ and $e^{2\nu}$ have finite values and remain positive throughout the stellar interior. At the center $e^{2\lambda(0)} = 1$ and $e^{2\nu(0)} = A^2 c_1^2 (-1)^{\frac{-3m-1}{2}}$. We show that in $r=0$ $\left(e^{2\lambda(r)}\right)'_{r=0} = \left(e^{2\nu(r)}\right)'_{r=0} = 0$ and this makes is possible to verify that the gravitational potentials are regular at the center. The energy density and radial pressure are positive and well behaved between the center and the surface of the star. In the center $\rho(r=0) = 3aC$ and $p_r(r=0) = 3maC$, therefore the energy density will be non-negative in $r=0$ and $p_r(r=0) > 0$. In the surface of the star $(r=R)$, $p_r(r=R) = 0$ and we have

$$R = \frac{\sqrt{2akC\left(k + \sqrt{k^2 - 12a^2 k}\right)}}{2akC} \tag{35}$$

For the radial pressure and density gradients we obtain

$$\frac{d\rho}{dr} = -2kC^2 r\left(1 - aCr^2\right) + 2akC^3 r^3 \tag{36}$$

$$\frac{dp_r}{dr} = mC\left[-2kCr\left(1 - aCr^2\right) + 2akC^2 r^3\right] \tag{37}$$

In order to maintain of causality, the radial sound speed should be within the limit $0 \leq v_{sr}^2 \leq 1$ in the interior of the star [80]. In this model, we have:

$$0 \leq v_{sr}^2 = \frac{dp_r}{d\rho} = \frac{1}{3} \leq 1 \tag{38}$$

On the boundary $r=R$, the solution must match the Reissner–Nordström exterior space–time as

$$ds^2 = -\left(1 - \frac{2M}{r} + \frac{Q^2}{r^2}\right)dt^2 + \left(1 - \frac{2M}{r} + \frac{Q^2}{r^2}\right)^{-1}dr^2 + r^2\left(d\theta^2 + \sin^2\theta d\varphi^2\right) \qquad (39)$$

and therefore, the continuity of $e^{2\lambda}$ and $e^{2\nu}$ across the boundary r=R is

$$e^{2\nu} = e^{2\lambda} = 1 - \frac{2M}{R} + \frac{Q^2}{R^2} \qquad (40)$$

Then for the matching conditions, we obtain

$$\frac{2M}{R} = aCR^2 + 2kC^2R^4 - 2akC^3R^6 \qquad (41)$$

With the quadratic regime and $n=2$, we have for the metric potentials $e^{2\lambda(0)} = 1$, $e^{2\nu(0)} = A^2 c_2^2(-1)^{\frac{-10Cam-1}{2}} e^{-\frac{F}{60}}$ and $\left(e^{2\lambda(r)}\right)'_{r=0} = \left(e^{2\nu(r)}\right)'_{r=0} = 0$ at the centre $r=0$. Again the gravitational potentials are regular in the origin. The energy density and radial pressure also are positive and well behaved in the stellar interior. In the center $\rho(r=0) = 6aC$ and $p_r = 36mCa^2$, therefore the energy density will be non-negative in $r=0$ and $p_r(r=0) > 0$. For the radial pressure and density gradients we obtain

$$\frac{d\rho}{dr} = C\left[-10a^2Cr - 2kCr\left(1 - aCr^2\right)^2 + 4akC^2r^3\left(1 - aCr^2\right)\right] \qquad (42)$$

$$\frac{dp_r}{dr} = 2mC\left[6a - 5a^2Cr^2 - kCr^2\left(1 - aCr^2\right)^2\right]\left[-10a^2Cr - 2kCr\left(1 - aCr^2\right)^2 + 4kC^2r^3a\left(1 - aCr^2\right)\right] \qquad (43)$$

The causality condition implies that

$$0 \leq 12amC - 10ma^2C^2r^2 - 2mkC^2r^2\left(1 - aCr^2\right)^2 \leq 1 \qquad (44)$$

On the boundary $r=R$, the solution must match the Reissner–Nordström exterior space–time and therefore for the matching conditions, we obtain:

$$\frac{2M}{R} = \frac{2aCR^2 + (2k - a^2)C^2R^4 - 4kaC^3R^6 + 2ka^2C^4R^8}{1 + aCR^2} \tag{45}$$

In Table 1 presents the values of the parameters chosen $a$, $K$, $m$ and $n$ for the linear and quadratic regimen. The mass and stellar radius obtained are also shown

**Table 1**. Parameters $m$, $a$, $k$, stellar radii and masses for linear and quadratic regime

| n | regime | m | a | k | M(M☉) | R(Km) |
|---|--------|---|---|---|-------|-------|
| 1 | linear | 1/3 | 0.003 | 0.00022 | 0.12M☉ | 5.1 |
| 2 | quadratic | 1 | 0.060 | 0.00022 | 2.05M☉ | 4.1 |

Where $M_\odot$ is the mass of the sun.

The figures 1, 2, 3, 4, 5, 6, 7, 8, 9 and 10 present the dependence of $\rho$, $\frac{d\rho}{dr}$, $e^{2\lambda}$, $e^{2\nu}$, $v_{sr}^2$, $p_r$, $\frac{dp_r}{dr}$, $M$, $\sigma^2$ and $\Delta$ with the radial coordinate for linear and quadratic regime, respectively. In all cases we have considered $C=1$.

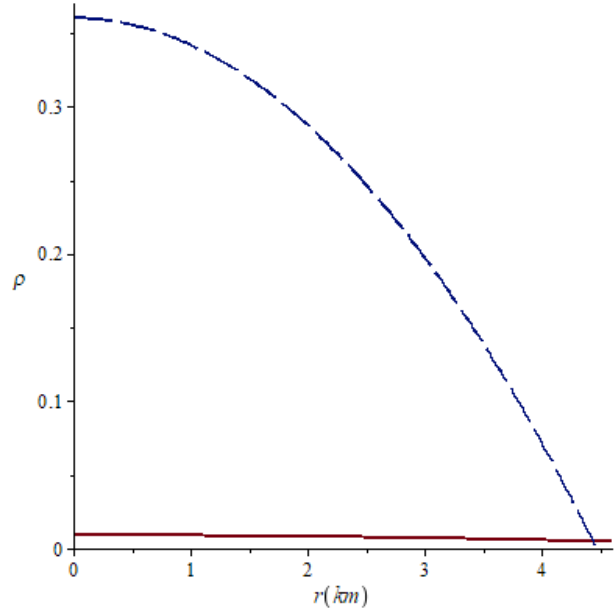

*Figure 1.* Energy density against radial coordinate for $n=1$ (solid line) and $n=2$ (long-dash line)

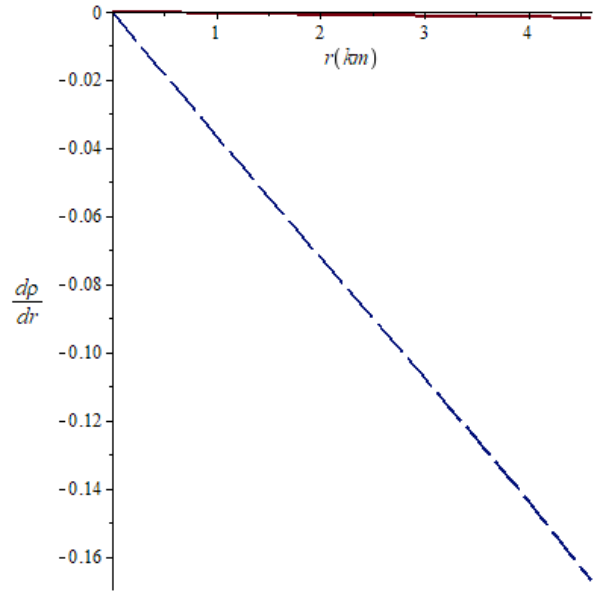

*Figure 2.* Density gradient against radial coordinate for *n*=1 (solid line) and *n*=2 (long-dash line)

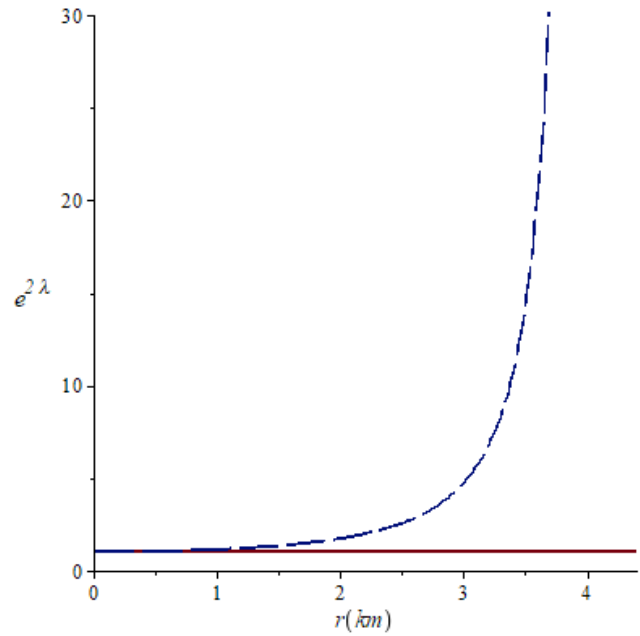

*Figure 3.* Metric potential $e^{2\lambda}$ against radial coordinate for *n*=1 (solid line) and *n*=2 (long-dash line)

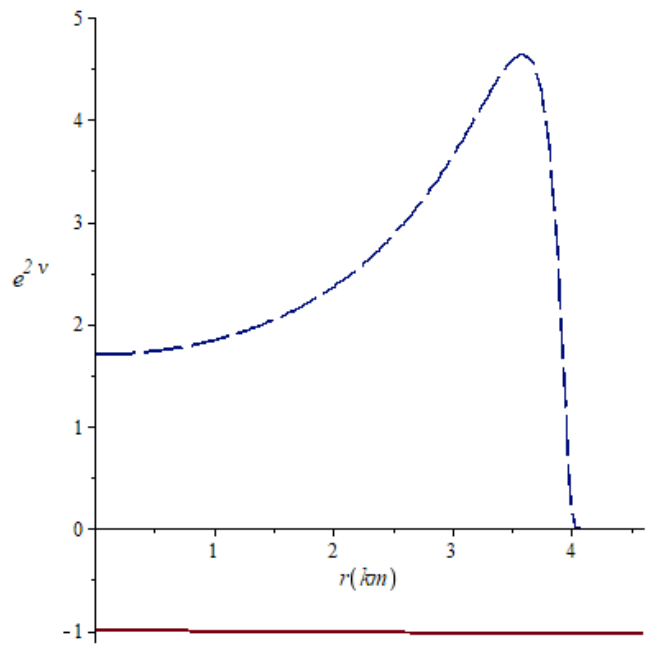

***Figure 4.*** Metric potential $e^{2\nu}$ against radial coordinate for $n=1$ (solid line) and $n=2$ (long-dash line)

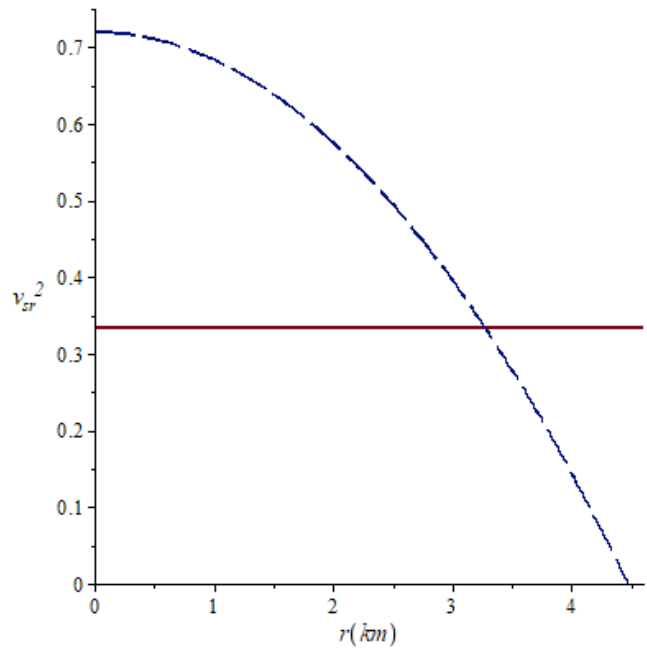

***Figure 5.*** Radial speed sound $v_{sr}^2$ against radial coordinate for $n=1$ (solid line) and $n=2$ (long-dash line)

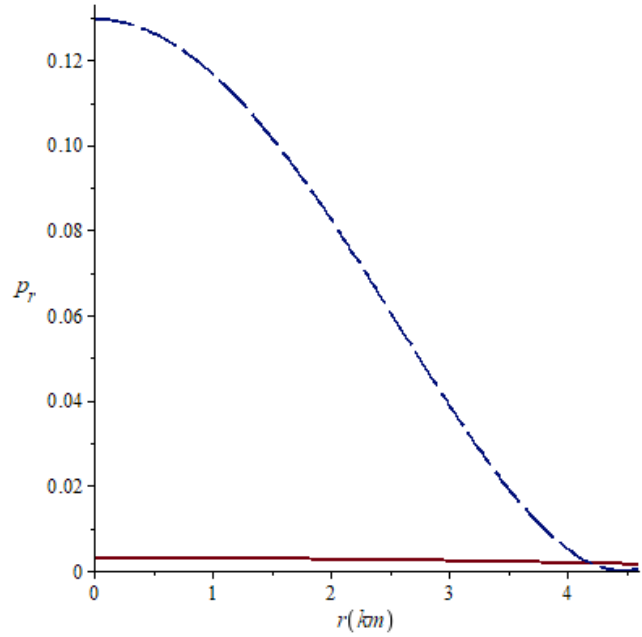

*Figure 6.* Radial pressure against radial coordinate for *n*=1 (solid line) and *n*=2 (long-dash line)

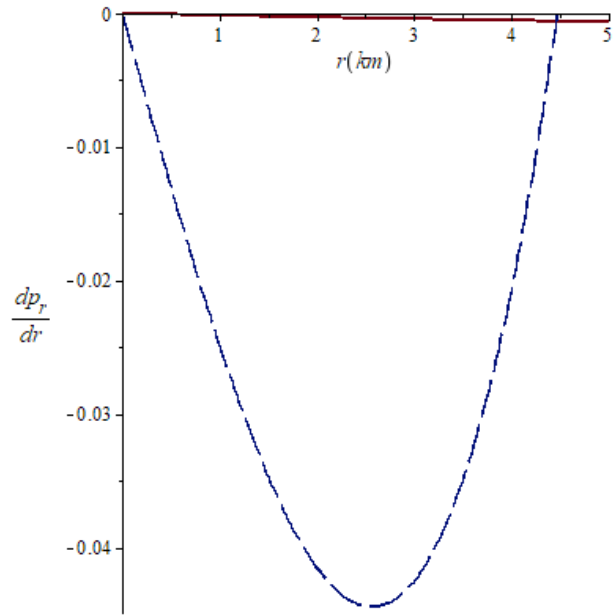

*Figure 7.* Radial pressure gradient against radial coordinate for *n*=1 (solid line) and *n*=2 (long-dash line)

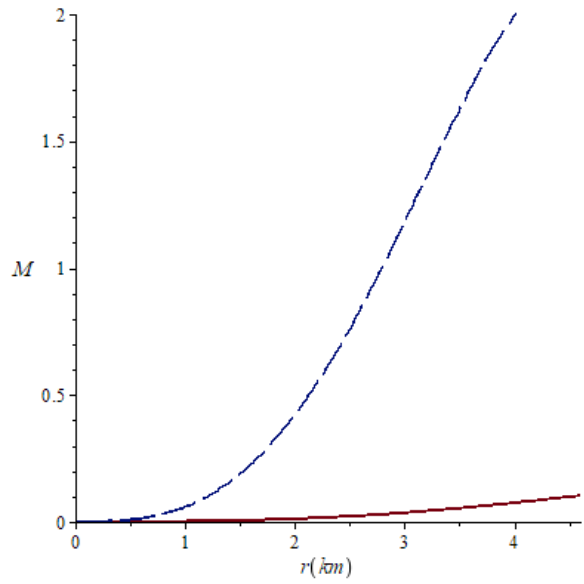

*Figure 8.* Mass function gradient against radial coordinate for *n*=1 (solid line) and *n*=2 (long-dash line)

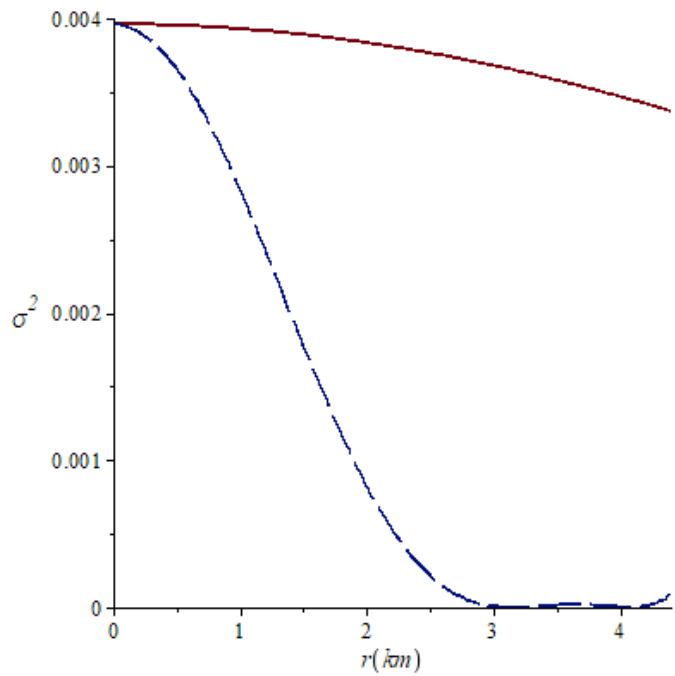

*Figure 9.* Charge density against radial coordinate for *n*=1 (solid line) and *n*=2 (long-dash line)

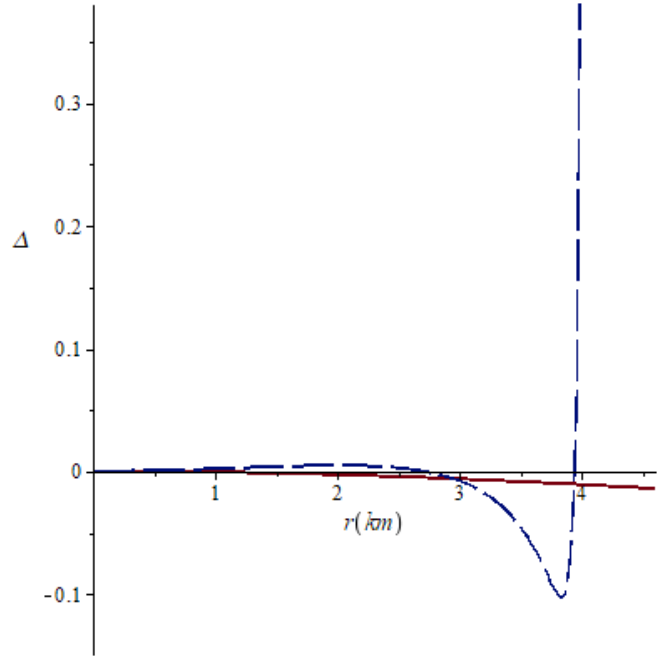

***Figure 10.*** Anisotropy against radial coordinate for *n=1* (solid line) and *n=2* (long-dash line)

In the Figure 1 is shown that the energy density remains positive, continuous and is monotonically decreasing function throughout the stellar interior for linear and quadratic regime. For *n=2*, the energy density takes higher values with the radial coordinate. In the Figure 2 it is noted that for the radial variation of energy density gradient $\frac{d\rho}{dr} < 0$ for the two cases studied, which is a condition for the physical acceptability of the model. In Figure 3 the metric potential $e^{2\lambda}$ in is continuously growing inside the star for *n=2* but for *n=1* not vary appreciably with the radial coordinate. In Figure 4 the metric potential $e^{2\nu}$ is continuous well behaved and shows an increase with the radial coordinate and then a decrease in the interior of the star for *n=2* but when *n=1* in linear regimen remains constant within the star. In Figure 5 the profile of radial speed sound is plotted for the linear and quadratic regime and is noted that $v_{sr}^2$ is always less that the unity and the causality condition is maintained in the stellar interior and it is important physical requirement as indicated Delgaty and Lake [80]. The radial pressure showed the same behaviour as energy density, shows the same behaviour as energy density, that is, growing within the star and vanishes at a greater radial distance, but takes higher values for *n=2* as shown in Figure 6. Again, according to Figure 7, the profile of $\frac{dp_r}{dr}$ shows that radial pressure is negative with the radial distance in linear and quadratic regime . In Figure 8, the mass function is regular, well behaved and strictly increasing from the centre to the surface of the star for *n*=1 and *n*=2. In Figure 9, that represent the variation of the charge density it is noted that is continuous, finite and is monotonically decreasing function for linear and quadratic regime. The anisotropy is plotted in figure 10 and it shows that vanishes at the centre of the star, i.e. *Δ(r=0)* =0. We can also that Δ admits higher values in quadratic regime for large values of radial distance.

We can compare the values calculated for the mass function wit observational data. For *n*=2 the values of *m*, *k* and *a* allow to obtain a mass of 2.05$M_\odot$ which can correspond to astronomic object PSR J1614-2230 [81] or it could also be associated to Cyg X-2 with a stellar mass of 1.78$M_\odot$ [82]. For the case *n*=1 we obtained comparable masses with the small star OGLE-TR-122b [83] with a mass of 0.09$M_\odot$ or the red dwarf Proxima Centauri whose mass is 0.123$M_\odot$ [84]. The values of the masses for these compact stars are tabulated is Table 2.

**Table 2**. The approximate values of the masses for the compact stars

| Compact Star | Masses $M(M_\odot)$ |
|---|---|
| PSR J1614-2230 | 1.908$M_\odot$ |
| Cyg X-2 | 1.78$M_\odot$ |
| OGLE-TR-122b | 0.123$M_\odot$ |

As stated earlier, we would point out that recent findings of the James Webb telescope discovered six galaxies that appeared between 500 and 700 million years after the Big Bang that contradicts the current understanding of how the first galaxies were formed [72]. Quantum picture with current modelling of astrophysics will provide a valuable tool in deeper knowledge of what is going on within these stellar objects, for example charge and energy densities, anisotropy, mass functions, and time transforms [66-69, 71]. Advantage of these theoretical to experimental design observables general conjectural modeling transforms will be its exceptional ability towards measurements by Instrumented PHYSICS techniques having algorithm that gives point-to-point parameters multiplicatively relating to signal to noise matrix measurable profile density. Point-to-point matrix intensity detection as well as measurement strategy provides a viable means of capturing observable measurable astrophysical signal/noise matrix of vibrational or sound and photonic or light gauge fields, equipped to detect point to point astrophysical light intensity signal/noise and
spectra density matrices of light signals [71].

## 6. Conclusion

In this paper we generated new exact models with the Thirukanesh-Ragel-Malaver ansatz for the metric potential considering linear and quadratic regime in presence of electric field. These models may be used in the description of compact objects charged and in the study of internal structure of strange star. We show that the new solutions are expressed in terms of polynomial and elementary functions which allow a physical study. A graphical analysis shows that the metric coefficients, radial pressure, energy density, charge density, mass function and anisotropy are regular at the origin and well behaved in the stellar interior. The new obtained models match smoothly with the Schwarzschild exterior metric across the boundary $r=R$ because matter variables and the gravitational potentials of this research are consistent with the physical analysis of these stars.

The new solutions can be related to stellar objects such as PSR J1614-2230, Cyg X-2 and OGLE-TR-122b. Physical features associated with the matter, radial pressure, density, anisotropy, charge density and metric potential and the plots generated suggests suggest that the model with $n=1$ similar to the red dwarf Proxima Centauri is well behaved [84]. It is expected that the results of this work can contribute to modeling of relativistic compact objects and configurations with anisotropic matter distribution. We have ansatz formalisms that connect astrophysics with the quantum nature of these anisotropic matter in stellar compact objects, with observable parameters derived from theoretical modeling to experimental measurements. These have all been necessitated by especially current findings of the James Webb Telescope of six earlier formed massive galaxies to peek into quantum nature with our newly developed point-to-point signal/noise matrix measurements of vibrational or sound and photonic or light gauge fields.